\documentclass[iop]{emulateapj}
  \usepackage{timesfonts}

\shorttitle{ASTEROSEISMOLOGY OF 16 CYG A \& B}
\shortauthors{METCALFE ET AL.}

\begin{document}

\newcommand{\CygAB}{16~Cyg~A~\&~B}
\newcommand{\CygA}{16~Cyg~A}
\newcommand{\CygB}{16~Cyg~B}
\newcommand{\Cyg}{16~Cyg}

%%%%%%%%%%%%%%%%%%%%%%%%%%%%%%%%%%%%%%%%%%%%%%%%%%%%%%%%%%%%%%%%%%%%%%%%%%

\title{Asteroseismic modeling of 16 Cyg A \& B using the complete Kepler 
data set}

\author{Travis S.\ Metcalfe$^1$, Orlagh L.\ Creevey$^{2,3}$, Guy R.\ 
Davies$^4$}

\affil{$^1$ Space Science Institute, 4750 Walnut St. Suite 205, Boulder CO 
80301 USA; travis@spacescience.org \\ $^2$ Laboratoire Lagrange, 
Universit\'e C\^ote d'Azur, Observatoire de la C\^ote d'Azur, CNRS, Blvd 
de l'Observatoire, CS 34229, 06304 Nice cedex 4, France \\ $^3$ Institut 
d'Astrophysique Spatiale, CNRS, UMR 8617, Universit\'e Paris XI, Batiment 
121, F-91405 Orsay Cedex, France \\ $^4$ School of Physics and Astronomy, 
University of Birmingham, Birmingham B15 2TT, UK}

\submitted{The Astrophysical Journal Letters, accepted}

\begin{abstract}

Asteroseismology of bright stars with well-determined properties from 
parallax measurements and interferometry can yield precise stellar ages 
and meaningful constraints on the composition. We substantiate this claim 
with an updated asteroseismic analysis of the solar-analog binary system 
\CygAB\ using the complete 30-month data sets from the {\it Kepler} space 
telescope. An analysis with the Asteroseismic Modeling Portal (AMP), using 
all of the available constraints to model each star independently, yields 
the same age ($t=7.0\pm0.3$~Gyr) and composition ($Z=0.021\pm0.002$, 
$Y_{\rm i}=0.25\pm0.01$) for both stars, as expected for a binary system. 
We quantify the accuracy of the derived stellar properties by conducting a 
similar analysis of a Kepler-like data set for the Sun, and we investigate 
how the reliability of asteroseismic inference changes when fewer 
observational constraints are available or when different fitting methods 
are employed. We find that our estimates of the initial helium mass 
fraction are probably biased low by 0.02--0.03 from neglecting diffusion 
and settling of heavy elements, and we identify changes to our fitting 
method as the likely source of small shifts from our initial results in 
2012. We conclude that in the best cases reliable stellar properties can 
be determined from asteroseismic analysis even without independent 
constraints on the radius and luminosity.

\end{abstract}


%%%%%%%%%%%%%%%%%%%%%%%%%%%%%%%%%%%%%%%%%%%%%%%%%%%%%%%%%%%%%%%%%%%%%%%%%%
\section{INTRODUCTION}\label{SEC1}

Asteroseismology is emerging as a powerful technique to determine the ages 
and other properties of field stars \citep{Metcalfe2014}, including the 
parent stars of planetary systems \citep{SilvaAguirre2015}. During the 
nominal {\it Kepler} mission, the data analysis and modeling methods for 
asteroseismic inference matured substantially, and the techniques are now 
being applied to brighter stars and members of well-characterized clusters 
for the {\it K2} mission \citep{Howell2014, Chaplin2015}. These studies 
will ensure that automated methods to determine stellar properties are as 
reliable as possible prior to the collection of asteroseismic data for 
bright stars all around the sky with the Transiting Exoplanet Survey 
Satellite \citep[TESS,][]{Ricker2014}.

The old solar analogs \CygAB\ are among the brightest stars ($V\sim6$) 
within the original {\it Kepler} field of view. Precise photometric 
measurements were obtained every 58.85 seconds almost continuously for 2.5 
years, between 2010 September 23 and 2013 April 8 (quarters Q7--Q16). 
\cite{Metcalfe2012} conducted an initial asteroseismic analysis of \CygAB\ 
using the first 3 months of {\it Kepler} data, identifying more than 40 
oscillation modes in each star. The optimal models, derived by fitting the 
observations of each star independently, had the same age and initial 
composition within the uncertainties---as expected for the components of a 
binary system. These initial results bolstered our confidence in the 
reliability of asteroseismic techniques.

Since the original study by \cite{Metcalfe2012} the time series for 
\CygAB\ has been extended by an order of magnitude, and radius constraints 
from interferometry are now available \citep{White2013}. The complete {\it 
Kepler} data sets have already been used to constrain the bulk helium 
abundances \citep{Verma2014}, and to determine rotation rates and 
inclinations \citep{Davies2015} for both components. Using these new 
observational constraints, and applying the improved modeling methods 
developed over the past several years, the aim of this Letter is to 
quantify the precision and accuracy of asteroseismic inferences for the 
bright targets that will be observed by future missions such as TESS and 
PLATO \citep{Rauer2014}.

In Section~\ref{SEC2} we outline the previously published observations 
that we adopt for our analysis, including a photometric data set for the 
Sun degraded to Kepler-like precision. We summarize recent updates to our 
methods in Section \ref{SEC3}, and we describe a new set of experiments 
using the Asteroseismic Modeling Portal \citep[AMP,][]{Woitaszek2009, 
Metcalfe2009} to probe the influence of different observational 
constraints and fitting methods. Finally, in Section~\ref{SEC4} we discuss 
the precision and accuracy of the derived stellar properties for \CygAB, 
using the solar results to identify any biases.

%%%%%%%%%%%%%%%%%%%%%%%%%%%%%%%%%%%%%%%%%%%%%%%%%%%%%%%%%%%%%%%%%%%%%%%%%%
\section{OBSERVATIONAL CONSTRAINTS}\label{SEC2}

To update our asteroseismic analysis, we adopted the sets of frequencies 
published by \cite{Davies2015}, including 54 and 56 oscillation modes for 
\CygAB\ respectively. These frequency sets were extracted from the power 
spectra of 928 days of short-cadence observations with a duty cycle of 
90.5 percent. Each set includes 15 consecutive orders for the radial 
($l=0$), dipole ($l=1$) and quadrupole ($l=2$) modes, as well as 9 and 11 
octupole ($l=3$) modes for \CygAB\ respectively. Compared to the 3 months 
of data presented in \cite{Metcalfe2012}, the longer time series analyzed 
by \cite{Davies2015} typically allowed the detection of 1--2 additional 
orders at lower and higher frequencies and improved the precision by a 
factor of 2--4 for previously detected modes, consistent with the 
expectations of \cite{Libbrecht1992}.

In addition to the frequencies, we adopted other observational constraints 
from spectroscopy, Hipparcos parallaxes \citep{vanLeeuwen2007}, and 
interferometry. The spectroscopic constraints and luminosities were 
similar to those used by \cite{Metcalfe2012}: $T_{{\rm eff},A} = 
5825\pm50$~K, [M/H]$_A = 0.10\pm0.09$, $T_{{\rm eff},B} = 5750\pm50$~K, 
[M/H]$_B = 0.05\pm0.06$ \citep{Ramirez2009}, $L_A = 1.56\pm0.05\ L_\odot$ 
and $L_B = 1.27\pm0.04\ L_\odot$ \citep{Metcalfe2012}. Note that we 
adopted 3$\sigma$ uncertainties on the spectroscopic metallicities to 
allow for potential systematic errors. Unlike the 2012 study, radius 
constraints are now available from the interferometric observations of 
\cite{White2013}, who used the CHARA array \citep{tenBrummelaar2005} with 
the PAVO beam combiner \citep{Ireland2008}. They found linear radii $R_A = 
1.22\pm0.02\ R_\odot$ and $R_B = 1.12\pm0.02\ R_\odot$, which were 
combined with an asteroseismic scaling relation \citep{Ulrich1986} to 
obtain mass estimates $M_A = 1.07\pm0.05\ M_\odot$ and $M_B = 1.05\pm0.04\ 
M_\odot$. The radius estimates are independent of asteroseismology, so we 
adopt them as constraints for the modeling presented in 
Section~\ref{SEC3}. We note the mass estimates here only for comparison 
with our final results (see discussion in Section~\ref{SEC4}).

To assess any biases in our modeling, we adopted a photometric data set 
for the Sun that was used by \cite{Davies2015} for validation of their 
analysis methods. The solar data were obtained from the red channel of the 
VIRGO instrument \citep{Frohlich1995} on the SoHO satellite 
\citep{Domingo1995} with white noise added to approximate the {\it Kepler} 
data for \CygAB. The solar power spectrum was analyzed using the same 
methods that were used for \Cyg, and the extracted frequencies were chosen 
to include the same set of modes (relative to the frequency of maximum 
power) detected in \CygA. The resulting solar frequency errors are 
comparable to those obtained for \CygA, and all other observational 
constraints for the Sun were fixed at the known values with the same 
fractional errors as \CygA: $T_{{\rm eff},\odot} = 5777\pm50$~K, 
[M/H]$_\odot = 0.00\pm0.09$, $L_\odot = 1.000\pm0.032$, and $R_\odot = 
1.000\pm0.016$.

%%%%%%%%%%%%%%%%%%%%%%%%%%%%%%%%%%%%%%%%%%%%%%%%%%%%%%%%%%%%%%%%%%%%%%%%%%
\section{ASTEROSEISMIC MODELING}\label{SEC3}

We used the observational constraints described in Section~\ref{SEC2} to 
obtain the optimal stellar properties with AMP from various data sets and 
fitting methods. The AMP science code uses a parallel genetic algorithm 
\citep{Metcalfe2003} to optimize the match between models produced by the 
Aarhus stellar evolution and pulsation codes \citep{jcd2008a,jcd2008b} and 
a given set of observations. The models are configured to use the OPAL 
2005 equation of state \citep{Rogers2002} with opacities from OPAL 
\citep{Iglesias1996} and \cite{Ferguson2005}, and nuclear reaction rates 
from the NACRE collaboration \citep{Angulo1999,Angulo2005}. Convection is 
treated using mixing-length theory \citep{BohmVitense1958} without 
overshoot, while diffusion and settling of helium is treated with the 
prescription of \cite{MichaudProffitt1993}. The AMP software has been in 
development since 2004, and below we briefly outline its history to place 
the current modeling approach in context.

\subsection{Updated physics and methods}

The AMP~1.0 software \citep{Metcalfe2009} reproduced the solar properties 
from Sun-as-a-star data, but we had more difficulty fitting the early 
asteroseismic observations from {\it Kepler} \citep{Metcalfe2010}. For 
AMP~1.1 \citep{Metcalfe2012, Mathur2012} we used the statistical errors on 
each oscillation frequency to assign weights, rather than using a 
combination of statistical and systematic errors that gave lower weight to 
higher frequencies where so-called ``surface effects'' dominated. We also 
split our $\chi^2$ quality metric into separate components for seismic and 
non-seismic observables, to prevent the many oscillation frequencies from 
overwhelming the relatively few and less precise spectroscopic 
constraints. For AMP~1.2 \citep{Metcalfe2014} we updated the model physics 
to use the NACRE reaction rates \citep{Angulo1999} instead of those from 
\cite{Bahcall1995} and the low-temperature opacities of 
\cite{Ferguson2005} instead of \cite{AlexanderFerguson1994}. We also began 
fitting two sets of frequency ratios $r_{010}$ and $r_{02}$ 
\citep{Roxburgh2003} in addition to the frequencies themselves, combining 
these three sets of observables with a set of spectroscopic and other 
constraints into an average $\chi^2$ quality metric.

We have made several important updates to AMP for the results presented 
here. First, we implemented the revised rate for the $^{14}N+p$ reaction 
determined by the NACRE collaboration \citep{Angulo2005}. Second, we now 
use only the frequency ratios $r_{010}$ and $r_{02}$ (along with $r_{13}$ 
for \CygAB\ where the $l=3$ modes are detected) rather than the 
individual frequencies for modeling. This allows us to avoid potential 
biases from applying an empirical correction for surface effects 
\citep[e.g.][]{Kjeldsen2008}. The frequency ratios are insensitive to the 
near-surface layers by design, so they allow a direct comparison between 
observations and models without resorting to any {\it ad hoc} frequency 
corrections. The frequency ratios are also insensitive to line-of-sight 
Doppler velocity shifts \citep[see][]{Davies2014}. Finally, we combine 
all of the observational constraints into a single $\chi^2$ quality 
metric to facilitate the statistical interpretation of the results. The 
average $\chi^2$ metric used in AMP~1.2 was designed to yield the optimal 
trade-off between complementary sets of constraints, but it complicated 
the determination of reliable uncertainties on the inferred stellar 
properties. We calculated the uncertainties in the same manner as 
described in \cite{Metcalfe2014}, but using the single $\chi^2$ value 
instead of an average $\chi^2$ to determine the likelihood of each 
sampled model. The AMP~1.2 fitting method can optionally still be used, 
and we have implemented the scaled solar surface correction proposed by 
\cite{jcd2012} to help minimize any resulting biases. We demonstrate 
below that our results are largely insensitive to the particular choice 
of fitting method.

% TABLE 1 ---------------------------------------------------------------- 
  \begin{deluxetable*}{lcccccccc}
  \tablecaption{Properties of the optimal models using various constraints and fitting methods\label{tab1}}
  \tablehead{\colhead{Case\hspace*{1.0in}} & \colhead{$R/R_\odot$} & \colhead{$M/M_\odot$} & \colhead{$L/L_\odot$} &
  \colhead{$t$/Gyr} & \colhead{$Z$} & \colhead{$Y_{\rm i}$} & \colhead{$\log g$} & \colhead{AMP\tablenotemark{a}}}
  \startdata
  \sidehead{\bf 16 Cyg A (Kepler)}
  1: all constraints\dotfill & $1.229\pm0.008$ & $1.08\pm0.02$ & $1.55\pm0.07$ & $7.07\pm0.26$ & $0.021\pm0.002$ & $0.25\pm0.01$ & $4.292\pm0.003$ & 767 \\
  2: without R\dotfill       & $1.225\pm0.008$ & $1.07\pm0.02$ & $1.53\pm0.07$ & $7.15\pm0.27$ & $0.020\pm0.002$ & $0.25\pm0.01$ & $4.291\pm0.002$ & 773 \\
  3: without R,L\dotfill     & $1.225\pm0.007$ & $1.07\pm0.02$ & $1.52\pm0.07$ & $7.12\pm0.23$ & $0.021\pm0.001$ & $0.25\pm0.01$ & $4.291\pm0.002$ & 776 \\
  4: fitting method\dotfill  & $1.229\pm0.007$ & $1.08\pm0.02$ & $1.55\pm0.06$ & $7.07\pm0.37$ & $0.021\pm0.002$ & $0.25\pm0.01$ & $4.292\pm0.002$ & 770 \\
  5: surface term\dotfill    & $1.229\pm0.008$ & $1.08\pm0.02$ & $1.55\pm0.07$ & $7.07\pm0.34$ & $0.021\pm0.002$ & $0.25\pm0.01$ & $4.292\pm0.002$ & 797 \\
  6: 3 months data\dotfill   & $1.229\pm0.008$ & $1.08\pm0.02$ & $1.55\pm0.07$ & $6.99\pm0.37$ & $0.024\pm0.002$ & $0.26\pm0.01$ & $4.292\pm0.003$ & 778 \\
  \sidehead{\bf 16 Cyg B (Kepler)}
  1: all constraints\dotfill & $1.116\pm0.006$ & $1.04\pm0.02$ & $1.25\pm0.05$ & $6.74\pm0.24$ & $0.022\pm0.003$ & $0.26\pm0.01$ & $4.359\pm0.002$ & 768 \\
  2: without R\dotfill       & $1.116\pm0.006$ & $1.04\pm0.02$ & $1.25\pm0.06$ & $6.74\pm0.23$ & $0.022\pm0.002$ & $0.26\pm0.01$ & $4.359\pm0.002$ & 774 \\
  3: without R,L\dotfill     & $1.116\pm0.006$ & $1.04\pm0.02$ & $1.24\pm0.06$ & $6.79\pm0.19$ & $0.022\pm0.002$ & $0.26\pm0.01$ & $4.359\pm0.002$ & 777 \\
  4: fitting method\dotfill  & $1.116\pm0.005$ & $1.04\pm0.01$ & $1.25\pm0.05$ & $6.89\pm0.28$ & $0.020\pm0.002$ & $0.25\pm0.01$ & $4.359\pm0.002$ & 771 \\
  5: surface term\dotfill    & $1.116\pm0.006$ & $1.04\pm0.02$ & $1.24\pm0.05$ & $6.96\pm0.29$ & $0.020\pm0.002$ & $0.25\pm0.01$ & $4.359\pm0.002$ & 798 \\
  6: 3 months data\dotfill   & $1.113\pm0.006$ & $1.03\pm0.02$ & $1.25\pm0.05$ & $6.66\pm0.27$ & $0.022\pm0.001$ & $0.27\pm0.01$ & $4.358\pm0.002$ & 779 \\
  \sidehead{\bf Sun (SoHO/VIRGO)}
  1: all constraints\dotfill & $1.003\pm0.006$ & $1.01\pm0.02$ & $0.98\pm0.04$ & $4.62\pm0.15$ & $0.018\pm0.002$ & $0.26\pm0.02$ & $4.439\pm0.002$ & 766 \\
  2: without R\dotfill       & $1.007\pm0.006$ & $1.02\pm0.02$ & $1.02\pm0.04$ & $4.53\pm0.15$ & $0.018\pm0.002$ & $0.26\pm0.01$ & $4.440\pm0.002$ & 772 \\
  3: without R,L\dotfill     & $1.010\pm0.005$ & $1.03\pm0.02$ & $1.00\pm0.04$ & $4.58\pm0.14$ & $0.018\pm0.001$ & $0.25\pm0.01$ & $4.442\pm0.002$ & 775 \\
  4: fitting method\dotfill  & $1.010\pm0.007$ & $1.03\pm0.02$ & $1.00\pm0.04$ & $4.52\pm0.18$ & $0.020\pm0.003$ & $0.25\pm0.01$ & $4.442\pm0.002$ & 769 \\
  5: surface term\dotfill    & $1.003\pm0.007$ & $1.01\pm0.02$ & $1.00\pm0.03$ & $4.61\pm0.18$ & $0.018\pm0.003$ & $0.26\pm0.02$ & $4.439\pm0.002$ & 799 \\
  6: 3 months data\dotfill   & $0.996\pm0.010$ & $0.99\pm0.03$ & $1.00\pm0.04$ & $4.77\pm0.18$ & $0.017\pm0.001$ & $0.27\pm0.02$ & $4.437\pm0.004$ & 800 \\
  \enddata
  \tablenotetext{a}{Comprehensive model output is available at 
  http://amp.phys.au.dk/browse/simulation/\#\#\#\\}
  \end{deluxetable*}
%-------------------------------------------------------------------------

\subsection{Results for \CygAB\ and the Sun}

We used AMP to investigate how the reliability of asteroseismic inference 
changes when various sets of observational constraints are adopted or when 
slightly different fitting methods are employed. The optimal asteroseismic 
properties of \CygAB\ and the Sun for each case are listed in 
Table~\ref{tab1}, and the variations are described below.

\begin{figure*}[t]
  \centerline{\includegraphics[width=3.375in]{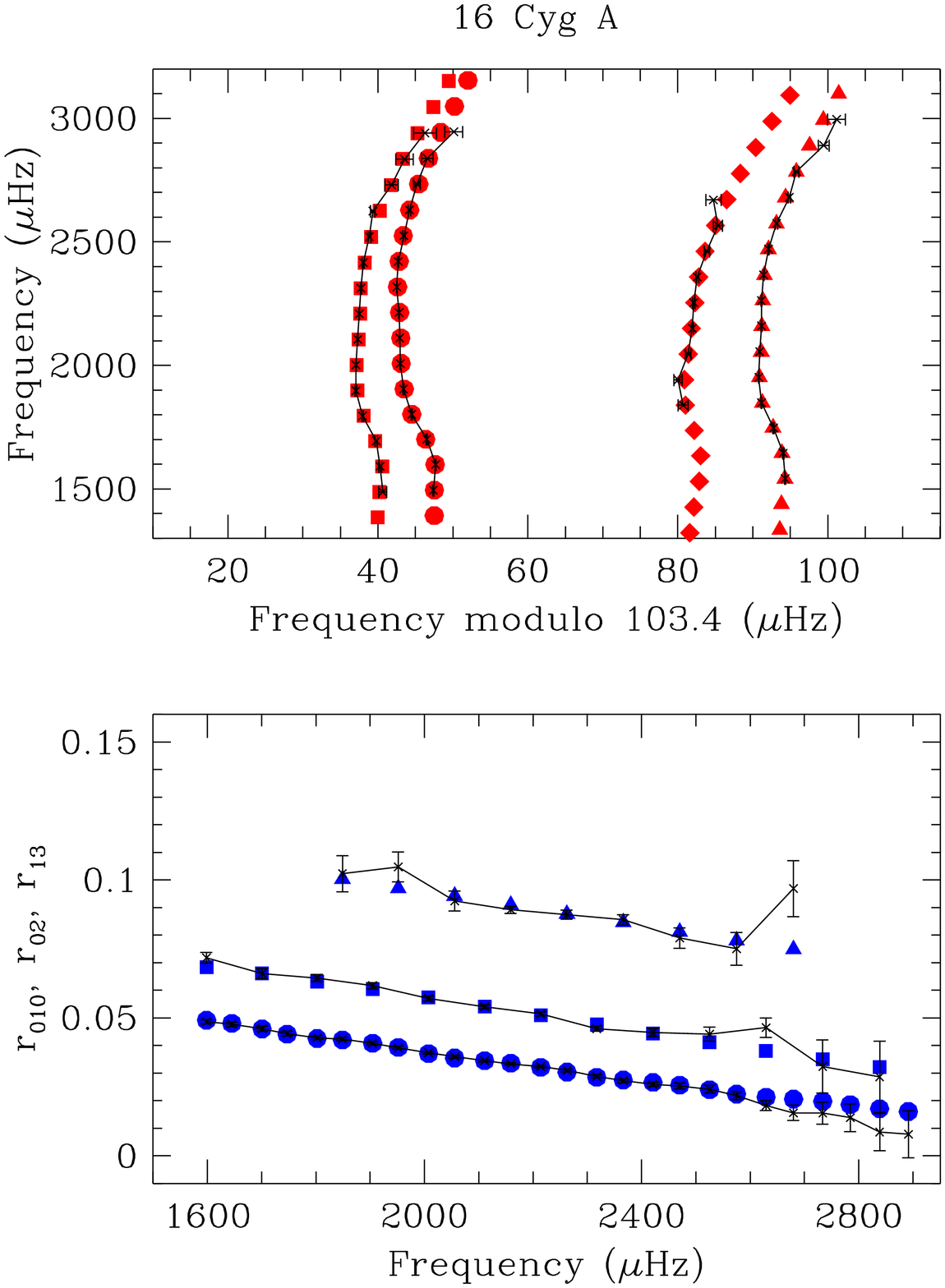}\hspace*{0.25in}\includegraphics[width=3.375in]{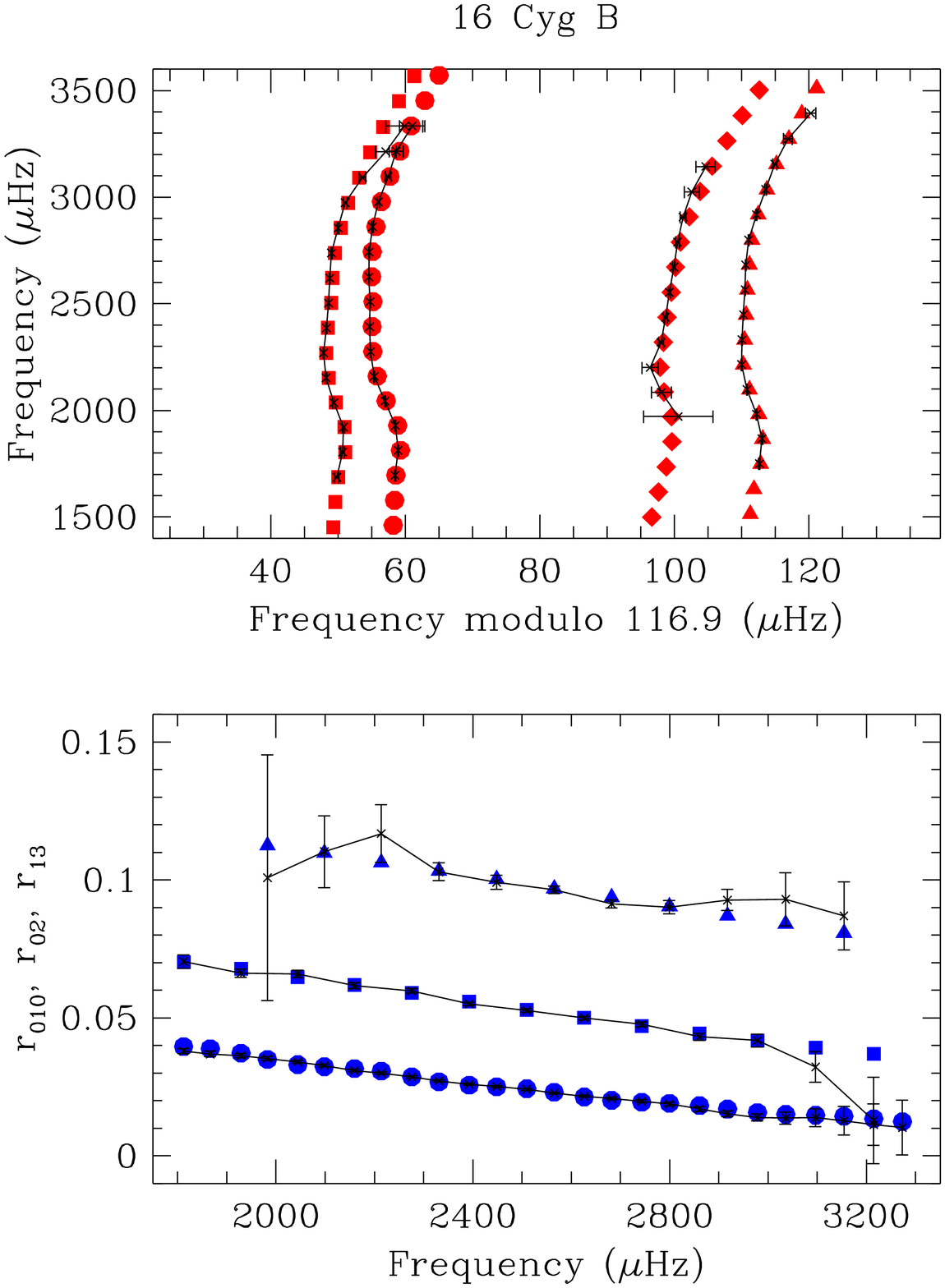}}
  \caption{\'Echelle diagrams~(top) and frequency ratios~(bottom) for \CygA\ 
  (left) and \CygB\ (right), with the observations shown as connected points 
  with errors. Top panels: frequencies of the optimal models from AMP are 
  shown using different symbols to indicate radial~(circles), 
  dipole~(triangles), quadrupole~(squares) and octupole~(diamonds) 
  oscillation modes. The scaled solar surface correction of \cite{jcd2012} 
  was applied to all model frequencies for illustration. Bottom panels: 
  frequency ratios $r_{010}$~(circles), $r_{02}$~(squares) and 
  $r_{13}$~(triangles) that were actually used as constraints for the 
  modeling.\label{fig1}\\}
  \end{figure*}

The most reliable properties, obtained by applying the updated fitting 
method described above to the complete set of observational 
constraints\footnote{The constraints used for each case are provided as an 
obs.dat file on the AMP website at \url{http://amp.phys.au.dk}.}, are 
shown in the first case labeled ``all constraints'' in Table~\ref{tab1} 
(AMP simulations 767, 768 and 766 for \CygA, \CygB\ and the Sun 
respectively). All three models match the spectroscopic and other 
constraints with a reduced $\chi^2 < 1$. The quality of the match to the 
asteroseismic constraints for \CygAB\ is illustrated in Figure~\ref{fig1} 
where the bottom panels show the match to the frequency ratios that were 
actually used as constraints for the modeling, while the top panels show 
the match to the frequencies after the scaled solar surface correction of 
\cite{jcd2012} has been applied to the raw model frequencies. All of the 
models that result from variations to the set of observational constraints 
and to the fitting methods yield comparable matches to both the 
asteroseismic and other constraints. The primary reason for exploring such 
variations is to quantify the absolute accuracy of the inferred stellar 
properties, and to assess the sensitivity of the results to details of the 
modeling strategy and data quality.

% FIGURE 1 --------------------------------------------------------------- 
%-------------------------------------------------------------------------

We can test the absolute accuracy of our results by eliminating the 
independent constraints on radius and luminosity to see whether the 
asteroseismic and spectroscopic constraints alone are sufficient to 
determine these properties. Interferometric radius constraints are 
available only for relatively nearby stars, so the second case (labeled 
``without R'' in Table~\ref{tab1}, AMP simulations 773, 774 and 772) 
removes this constraint from the $\chi^2$ value that guides the search. 
For \CygA\ and the Sun this leads to small shifts within the quoted 
uncertainties for the optimal stellar properties, while only the 
uncertainties change for \CygB. Luminosity constraints are currently 
available for a small fraction of stars in the {\it Kepler} field, at 
least until parallaxes are released by the {\it Gaia} mission 
\citep{Perryman2001}. Parallaxes are also needed to convert the angular 
measurements from interferometry into a linear radius, so our third case 
(labeled ``without R,L'' in Table~\ref{tab1}, AMP simulations 776, 777 and 
775) omits both the radius and luminosity constraints. Again there are 
marginal shifts in the optimal stellar properties, insignificant for 
\CygAB\ but slightly larger than the uncertainties for the solar radius 
and mass. In general, the reliability of asteroseismic inference seems to 
degrade more significantly for the Sun than for \CygAB\ when the radius 
and luminosity constraints are not available.

Large samples of {\it Kepler} stars have previously been fit using 
slightly different methods to assess the quality of the match between the 
models and observations. To evaluate the impact of these differences on 
the inferred stellar properties, we fit the full set of constraints from 
the first case using two alternative modeling strategies. For the fourth 
case (labeled ``fitting method'' in Table~\ref{tab1}, AMP simulations 770, 
771 and 769) we used the AMP~1.2 fitting method \citep{Metcalfe2014}, 
which calculates an average $\chi^2$ from four sets of observables 
including the frequencies corrected for surface effects with a 
solar-calibrated power law following \cite{Kjeldsen2008}. Comparing the 
inferred stellar properties to the first case, \CygA\ yields slightly 
different uncertainties, \CygB\ is shifted within the uncertainties, and 
only the Sun shows marginally significant shifts in radius and mass. To 
probe the possible source of these biases, for the fifth case (labeled 
``surface term'' in Table~\ref{tab1}, AMP simulations 797, 798 and 799) we 
repeated the fourth case using the scaled solar surface correction of 
\cite{jcd2012} to replace the \cite{Kjeldsen2008} prescription. The 
differences were insignificant for \CygAB, but for the Sun the alternative 
surface correction effectively eliminated the biases.

Finally, we examine the influence of using the longer asteroseismic time 
series by applying our updated fitting method to the original data sets 
for \CygAB\ published in \cite{Metcalfe2012}, along with a comparable 
3-month solar data set from VIRGO. This final case (labeled ``3 months 
data'' in Table~\ref{tab1}, AMP simulations 778, 779 and 800) should be 
compared to the second case ``without R'', since the original data sets 
did not include the radius constraints of \cite{White2013}. For \CygAB\ 
these two cases yield similar stellar properties, indicating that 
differences relative to the \cite{Metcalfe2012} results can be attributed 
mostly to the updated model physics rather than the extended time series. 
For the Sun, the properties inferred from the shorter time series 
reproduce the solar radius, mass and luminosity more accurately. This 
seems to be a consequence of the fewer and lower-precision asteroseismic 
constraints, which gives the spectroscopic and other constraints more 
relative weight when all of the observations are combined into a single 
$\chi^2$ quality metric.

%%%%%%%%%%%%%%%%%%%%%%%%%%%%%%%%%%%%%%%%%%%%%%%%%%%%%%%%%%%%%%%%%%%%%%%%%%
\section{DISCUSSION}\label{SEC4}

We have determined precise asteroseismic properties for the solar-analog 
binary system \CygAB, and we have quantified the accuracy of the results 
by varying the input constraints and fitting methods, and by applying the 
same techniques to the Sun. We obtain a radius precision of 0.6\%, a mass 
precision of 2\%, and an age precision of 3\%. Our results for the Sun 
suggest that the systematic errors are smaller than or comparable to the 
statistical uncertainties, consistent with the conclusions of previous 
studies \citep[e.g.][]{Lebreton2014, SilvaAguirre2015}. The radii and 
masses of \CygAB\ are consistent with the estimates of \cite{White2013} 
from interferometry and scaling relations, but systematically lower than 
the original AMP results in \cite{Metcalfe2012}. These differences 
persist when the radius constraint is omitted, or when the 2012 data set 
is analyzed with updated methods, suggesting that the use of frequency 
ratios as asteroseismic constraints is responsible. The luminosities of 
\CygAB\ agree with the values derived by \cite{Metcalfe2012} from the 
Hipparcos parallaxes, though the best models for both components have 
luminosities slightly below the constraints. The solar radius, mass and 
luminosity are all recovered faithfully within the quoted precision, 
including small biases of +0.3\% in radius, +1\% in mass and $-$2\% in 
luminosity. Note that results for significantly more massive or more 
evolved stars may not be as precise or accurate, even with comparable 
data quality.

Although we fit the observations of \CygAB\ individually, the results for 
the two stars have a common age ($t=7.0\pm0.3$~Gyr) and composition 
($Z=0.021\pm0.002$, $Y_{\rm i}=0.25\pm0.01$) within the uncertainties, as 
expected for the components of a binary system. The updated age is 
slightly older and more precise than the result presented in 
\cite{Metcalfe2012}, and the updated methods in AMP now yield a consistent 
age for both components. The closest agreement between the two sets of 
models is obtained when using the AMP~1.2 fitting method with the scaled 
solar surface correction of \cite{jcd2012}. The comparable fits to solar 
data agree with the seismic age of the Sun ($4.60\pm0.04$~Gyr) as 
determined by \cite{Houdek2011}, with the ensemble of models in 
Table~\ref{tab1} showing a total age spread of only 0.25~Gyr (5\%).

As with the original results of \cite{Metcalfe2012}, we find values for 
the initial helium mass fraction that are systematically low compared to 
other estimates. \cite{Verma2014} used deviations from uniform frequency 
spacing in the oscillation modes to constrain the acoustic depths of the 
base of the convection zone (BCZ, $\tau_{cz}$) and the helium ionization 
zone. The amplitude of deviations due to the latter source can be 
calibrated with models to determine the current helium mass fraction in 
the stellar envelope. The weighted average of three different methods 
using two different models for calibration yields $Y_{\rm 
s}=0.241\pm0.004$ for \CygA\ and $Y_{\rm s}=0.244\pm0.004$ for \CygB. In 
our models $Y_{\rm s}$ is systematically lower by 0.02--0.03, confirming 
that we are biased toward low initial helium. This issue had previously 
been identified by \cite{Gruberbauer2013} from Bayesian modeling of the 
2012 data set for \CygAB, and our results for the Sun show the same 
problem. We speculate that the bias toward low initial helium may be due 
to neglecting diffusion and settling of heavy elements, which is not 
stable under all conditions for the models used in AMP. We hope to remedy 
this issue with future development of AMP~2.0, which will use the MESA 
stellar evolution code \citep{Paxton2011, Paxton2013, Paxton2015} and the 
GYRE pulsation code \citep{Townsend2013}.

An additional diagnostic of our best models is whether they agree with the 
convection zone depths derived by \cite{Verma2014}. Although unrealistic 
surface boundary conditions in the models hinder direct comparisons, 
\cite{Mazumdar2014} suggest comparing the {\it fractional} acoustic radius 
of the BCZ from observations and models. The total acoustic radius of the 
star $T_0$ can be estimated from the mean frequency spacing of radial 
modes $\Delta_0$ using $T_0 \sim 1/(2\Delta_0)$. The fractional acoustic 
radius of the BCZ is then $T_{\rm BCZ}/T_0 = 1-(\tau_{cz}/T_0)$. The 
comparable quantity for a model is obtained from a numerical integration 
of the inverse sound speed from the center of the model to the BCZ, 
relative to a full integration out to the photosphere. A weighted average 
of the values of $T_{\rm BCZ}/T_0$ from three methods using the data in 
Tables 1 and 3 of \cite{Verma2014} gives $0.368\pm0.012$ for \CygA\ and 
$0.391\pm0.028$ for \CygB, while our best models yield 0.386 and 0.381 
respectively---slightly above the constraint for \CygA\ ($+1.5\sigma$), 
but consistent for \CygB\ ($-0.4\sigma$). Future efforts should consider 
how to include the fractional acoustic radius of interior features as 
additional constraints for the modeling.

Automated asteroseismic modeling has advanced significantly in the past 
few years, and we are on track to take full advantage of the large data 
sets that will emerge from future missions like TESS (1--12 month 
time-series) and PLATO (5--36 months per field). The analysis presented 
in this Letter demonstrates that high precision and reasonable accuracy 
is possible for the brightest asteroseismic targets, with or without 
independent constraints on the radius and luminosity. It is also clear 
that our results are largely insensitive to fine details of the modeling 
strategy, but significant biases are possible when using simple power law 
corrections for surface effects and when neglecting diffusion and 
settling of heavy elements. Efforts are underway to define the character 
of these systematic errors across a broad range of stellar properties, 
and to minimize or correct the model deficiencies that contribute to the 
biases. We look forward to additional opportunities to demonstrate the 
reliability of asteroseismic properties using observations of cluster 
members, binary systems, and interferometric targets.

%%%%%%%%%%%%%%%%%%%%%%%%%%%%%%%%%%%%%%%%%%%%%%%%%%%%%%%%%%%%%%%%%%%%%%%%%% 
\acknowledgments This work was supported by NASA grants NNX13AE91G and 
NNX15AF13G. Computational time at the Texas Advanced Computing Center was 
provided through XSEDE allocation TG-AST090107. G.R.D.\ acknowledges 
support from the UK Science and Technology Facilities Council (STFC).

%%%%%%%%%%%%%%%%%%%%%%%%%%%%%%%%%%%%%%%%%%%%%%%%%%%%%%%%%%%%%%%%%%%%%%%%%% 

% FIGURE 1 --------------------------------------------------------------- 
%-------------------------------------------------------------------------

% TABLE 1 ---------------------------------------------------------------- 
%-------------------------------------------------------------------------

\end{document}